\documentclass[3p,times,procedia]{elsarticle}
\usepackage{nupha_ecrc}

\volume{00}

\firstpage{1}

\journalname{Nuclear Physics A}

\runauth{}

\jid{nupha}

\jnltitlelogo{Nuclear Physics A}

\usepackage{amssymb}

 \biboptions{square,sort&compress}

\usepackage[figuresright]{rotating}
\usepackage{lineno}

\begin{document}

\begin{frontmatter}



\dochead{XXVIIIth International Conference on Ultrarelativistic Nucleus-Nucleus Collisions\\ (Quark Matter 2019)}

\title{Jet quenching and acoplanarity via hadron-jet measurements in pp and Pb--Pb collisions at 5.02 TeV with ALICE}


\author{Yaxian MAO\corref{col1}}
\author{for the ALICE Collaboration}

\address{Key Laboratory of Quark $\&$ Lepton Physics of Ministry of Education, Institute of Particle Phisics, Central China Normal University, \\ Wuhan 430079, China}

\cortext[col1] {Supported by the National Science Foundation of China (Grant No. 11975005 and 11505072).}

\begin{abstract}
Measurements of the semi-inclusive distribution of charged jets recoiling from 
a trigger hadron in pp and Pb--Pb collisions at $\sqrt{s_{\mathrm{NN}}}$ = 5.02 TeV are presented. This 
technique provides precise, data-driven subtraction of the large uncorrelated background 
in jet measurements. It uniquely enables the exploration of medium-induced modification 
of jet production and acoplanarity over wide phase space, including low transverse momentum ($p_\mathrm{T}$) 
and large resolution parameter ($R$) jets. This proceeding reports the measurements of medium-induced 
jet energy redistribution through the comparison of trigger-normalized recoil jet yields 
in pp and Pb--Pb collisions, and for jets with different 
$R$. 
\end{abstract}


\end{frontmatter}


\section{Introduction}
\label{intro}
The experimental heavy-ion program aims to explore the phases of nuclear matter and the properties of the Quark Gluon Plasma (QGP), the primordial state of matter that existed during few hundreds of microseconds in the early Universe until the quark-hadron phase transition. The hot and dense de-confined matter, QGP, is formed in collisions of heavy ions at ultra-relativistic energies~\cite{Bjorken}. The study of the QGP in the laboratory allows us to improve our understanding of quantum chromodynamics (QCD), the theory of the strong interaction. 

Hard probes (high transverse momentum $p_\mathrm{T}$ or high mass), that involve scales smaller than the temperature of the thermal medium, are exploited as “external” probes of the QGP to understand its interactions depending on the color charge, mass and flavor of the probe. Jet quenching is a phenomenon that has been observed for all hadron species at high $p_\mathrm{T}$ as a pronounced suppression of the production in heavy-ion collisions relative to pp collisions~\cite{PHENIX,STAR,CMS_Raa2012,ATLAS,ALICE}. These high-$p_\mathrm{T}$ hadrons originate from hard scattered partons produced in the initial stage of the collisions that lost energy while traversing the medium. The partonic shower generated after the hard scattering process by subsequent fragmentation is measured as a collimated jet of hadrons. Due to the jet quenching, both the jet energy and its distribution can be modified by the medium. Comparison of the jet production between proton-proton and heavy-ion collisions reveals these modifications experimentally. 
The measurement of reconstructed jets over a wide range in jet energy and different jet resolution parameters ($R$) is required for comprehensive understanding of jet quenching in heavy-ion collisions~\cite{ppjetALICE}. 
However, such measurements are challenging because of the presence of complex, uncorrelated background to the jet signal, and the need to minimize biases in the selected jet population imposed by
background subtraction techniques. Multiple, complementary measurement approaches are therefore needed to elucidate the jet quenching mechanism using reconstructed jets.

In this proceeding, we report the measurements of the semi-inclusive distribution of charged jets recoiling from a trigger hadron in pp and Pb--Pb collisions at $\sqrt{s_{\mathrm{NN}}} = 5.02$ TeV, using the same approach as developed in central Pb--Pb collisions at $\sqrt{s_{\mathrm{NN}}} = 2.76$ TeV~\cite{ALICE_hJet}. This 
technique provides precise, data-driven subtraction of the large uncorrelated background in jet measurements. It enables the exploration of medium-induced modification of jet production and acoplanarity over a wide phase space, including low $p_\mathrm{T}$ for large jet resolution parameter $R$. 

\section{Analysis}
\label{ana} 
The Pb--Pb collision data were recorded during the 2018 LHC heavy-ion run
at $\sqrt{s_{\mathrm {NN}}} = 5.02$ TeV. The pp collision data used to provide the reference study
were recorded by ALICE during the 2017 pp data taking at $\sqrt{s} = 5.02$
TeV, using a minimum bias trigger. A detailed description of the trigger determination in ALICE can be found in ~\cite{detector}.

The selection of events containing a hard process
(``hard-process selection'') is based on the presence of a
high-$p_{\mathrm{T}}$ hadron trigger. Hadrons with $p_{\mathrm{T}}$ larger than about 5~$\mathrm{GeV/}c$ are expected to originate
primarily from fragmentation of jets.
The analysis is based on the semi-inclusive distribution of charged jets
recoiling from a high-$p_{\mathrm{T}}$ charged hadron trigger, normalized by the number of trigger hadrons, with the trigger hadron selected
within a limited $p_{\mathrm{T,trig}}$ interval (Trigger Track, or TT, class). 
Trigger hadrons within $|\eta|<0.9$ and $5<p_{\mathrm{T,trig}}<7~ \mathrm{GeV/}c$ are referred as the Reference TT class, denoted by TT(5,7). Those within $|\eta|<0.9$ and $20<p_{\mathrm{T,trig}}<50~ \mathrm{GeV/}c$ are referred as 
the Signal TT class, denoted by TT(20,50). 
The number of trigger hadrons ($N_\mathrm{trig}$) is used to normalize the recoil jets measured in the selected TT intervals.  

Jets are reconstructed using charged particle tracks with the
$k_\mathrm{T}$~\cite{Cacciari:2011ma} and anti-$k_\mathrm{T}$ algorithms~\cite{FastJetAntikt}, with infrared cut off for tracks $p_{\mathrm{T}} > 0.15~ \mathrm{GeV/}c$ in the acceptance of $|\eta_\mathrm{jet}| < 0.9 - R $. Uncorrelated background to the recoil jet signal is corrected at the level of ensemble-averaged distributions, without
event-by-event discrimination of jet signal from background, using a technique that exploits the phenomenology of jet production in QCD~\cite{ALICE_hJet}. 
The observable, $\Delta_\mathrm{recoil}$, is the difference between two semi-inclusive recoil jet
distributions, for the Signal and Reference TT classes \cite{deBarros:2012ws}. \begin{equation}
\Delta_\mathrm{recoil} = 
\frac{1}{N_\mathrm{trig}}\frac{\mathrm{d}^{2}N_\mathrm{jet}}{\mathrm{d}p_{\mathrm{T,jet}}^\mathrm{ch}\mathrm{d}\eta_\mathrm{jet}}\Bigg\vert_{p_{\mathrm{T,trig}}\in{\mathrm{TT}_{\mathrm{Sig}}} }-
c_\mathrm{Ref}\cdot \frac{1}{N_\mathrm{trig}}\frac{\mathrm{d}^{2}N_\mathrm{jet}}{\mathrm{d}p_{\mathrm{T,jet}}^\mathrm{ch}\mathrm{d}\eta_\mathrm{jet}}\Bigg\vert_{p_{\mathrm{T,trig}}\in{\mathrm{TT}_{\mathrm{Ref}}}}
\label{eq:DRecoil}
\end{equation}

\noindent
The scale factor $c_\mathrm{Ref}$, which accounts for the combined effects of invariance of total jet yield with trigger hadron $p_\mathrm{T}$,
 is within a few percent of unity~\cite{ALICE_hJet}. 
Using $\Delta_\mathrm{recoil}$, the uncorrelated jet yield is suppressed in a purely data-driven way. Therefore, the $\Delta_\mathrm{recoil}$ distribution of jet yield is clean of combinational background. This is important for acoplanarity measurements to suppress the effects of Multiple Parton Interactions (MPI) (see Ref.~\cite{ALICE_hJet}). 
The raw $\Delta_\mathrm{recoil}$ distribution must be corrected for smearing of recoil jet energy by fluctuations of energy density in the underlying event. The correction is carried out using an unfolding technique~\cite{DAgostini:1994zf}, as implemented in the RooUnfold package. 
After corrections, $\Delta_\mathrm{recoil}$ represents the change in the distribution of jets recoiling against a trigger hadron, as the trigger hadron $p_{\mathrm{T}}$ changes from the Reference to Signal TT interval. 
The differential recoil jet distributions as a function of $p_{\mathrm{T,jet}}^\mathrm{ch}$, are reported for $R$ = 0.2, 0.4 and 0.5, over the range
$20< p_{\mathrm{T,jet}}^\mathrm{ch} <100~ \mathrm{GeV/}c$. 

\section{Results}
\label{result} 
\begin{figure*}[htbp]
	\begin{center}
	\includegraphics[width=1.0\textwidth]{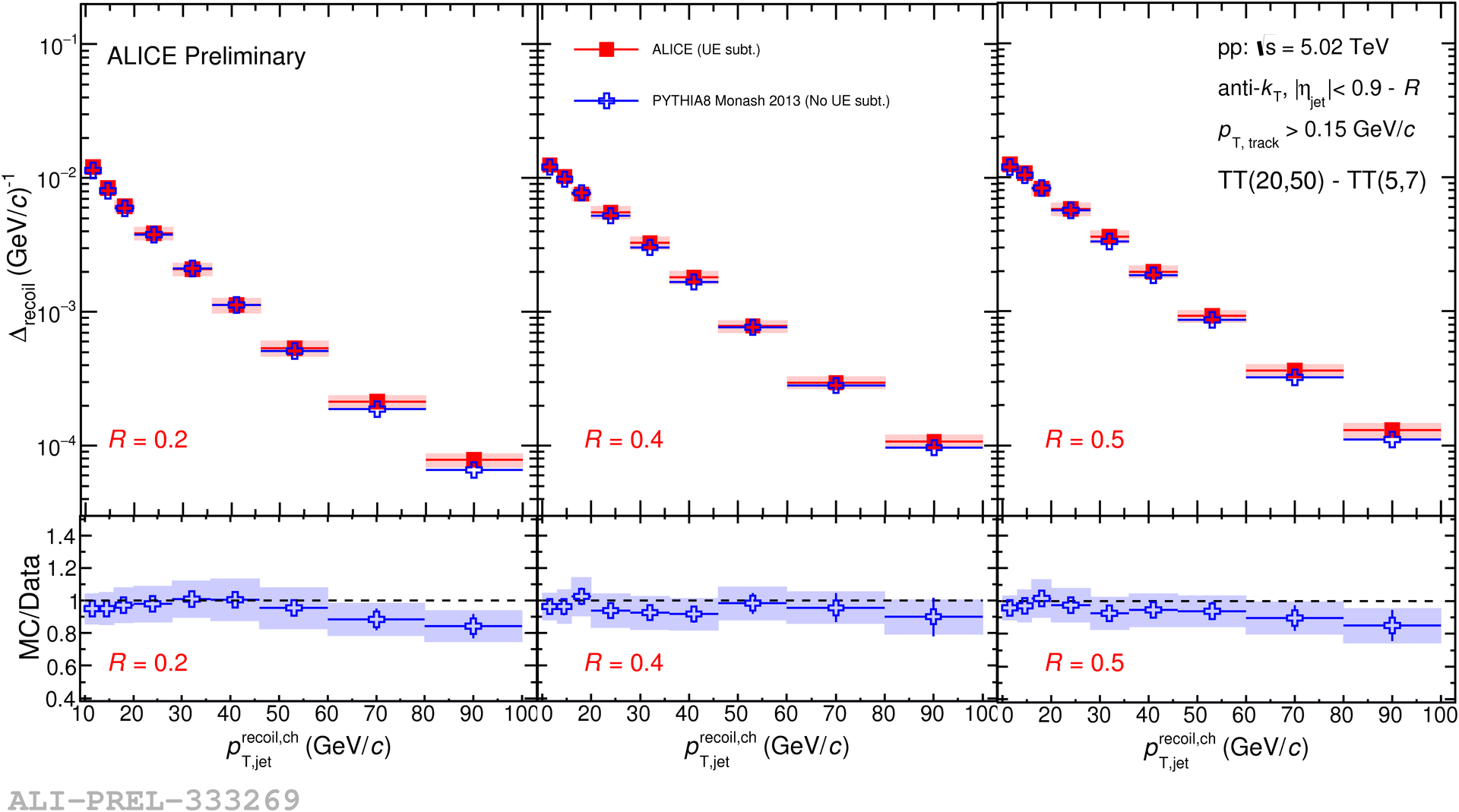}	
	\end{center}
	\caption[width=0.8\textwidth]{Comparison of $\Delta_\mathrm{recoil}$ distribution in pp collisions and PYTHIA distribution for $R$ = 0.2, 0.4 and 0.5 using TT(20,50) and TT(5,7) intervals. }
	\label{fig:ppRecoil}
\end{figure*}

Figure~\ref{fig:ppRecoil} shows the corrected $\Delta_\mathrm{recoil}$ distributions for jet resolution parameter $R$ = 0.2 (left), 0.4 (middle) and 0.5 (right) and a comparison with the PYTHIA8 MC~\cite{Pythia8} predictions, the bottom panels show the ratio between MC and data.  

\begin{figure*}[htbp]
	\begin{center}
	\includegraphics[width=0.325\textwidth]{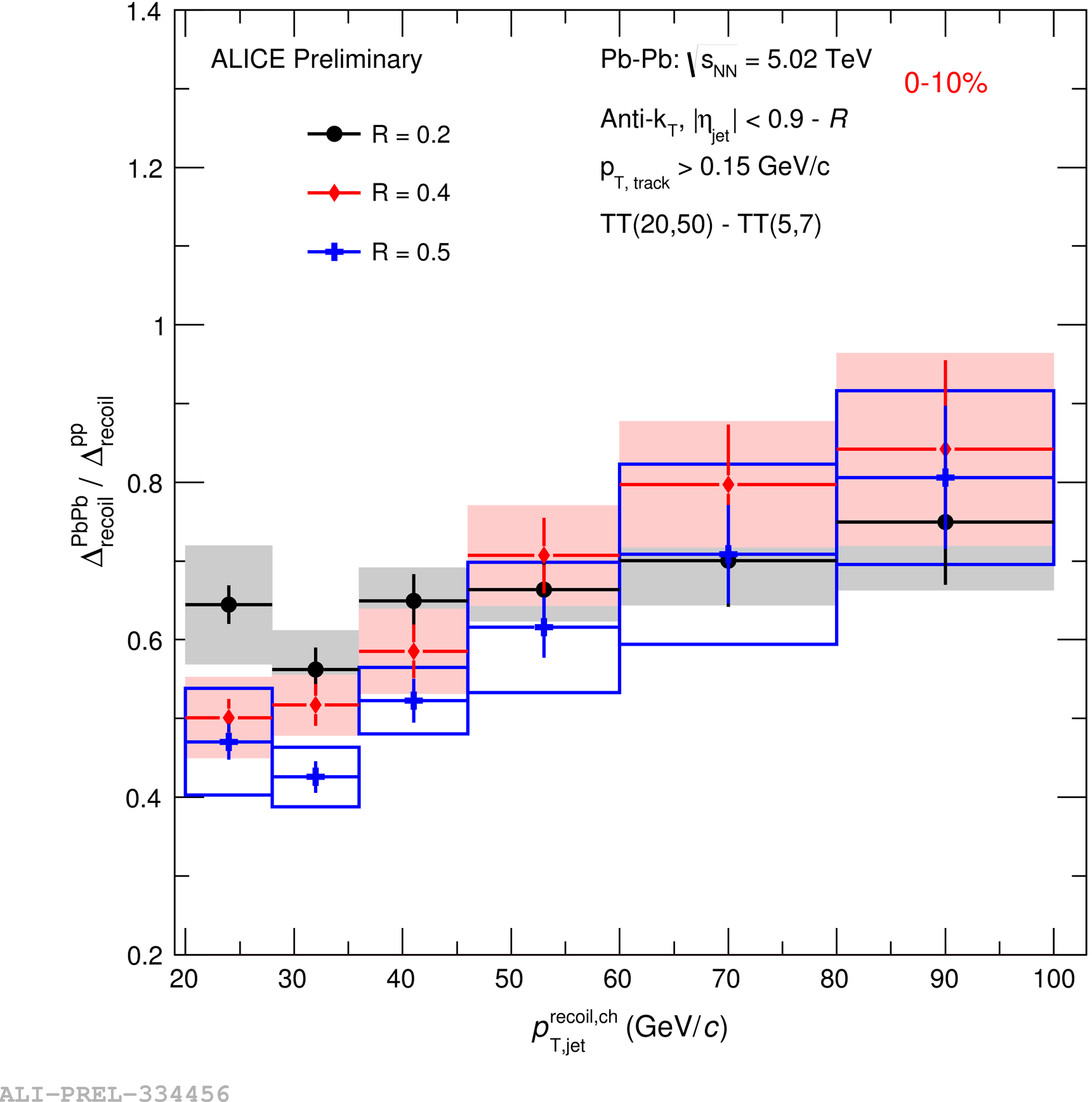}	
	\includegraphics[width=0.325\textwidth]{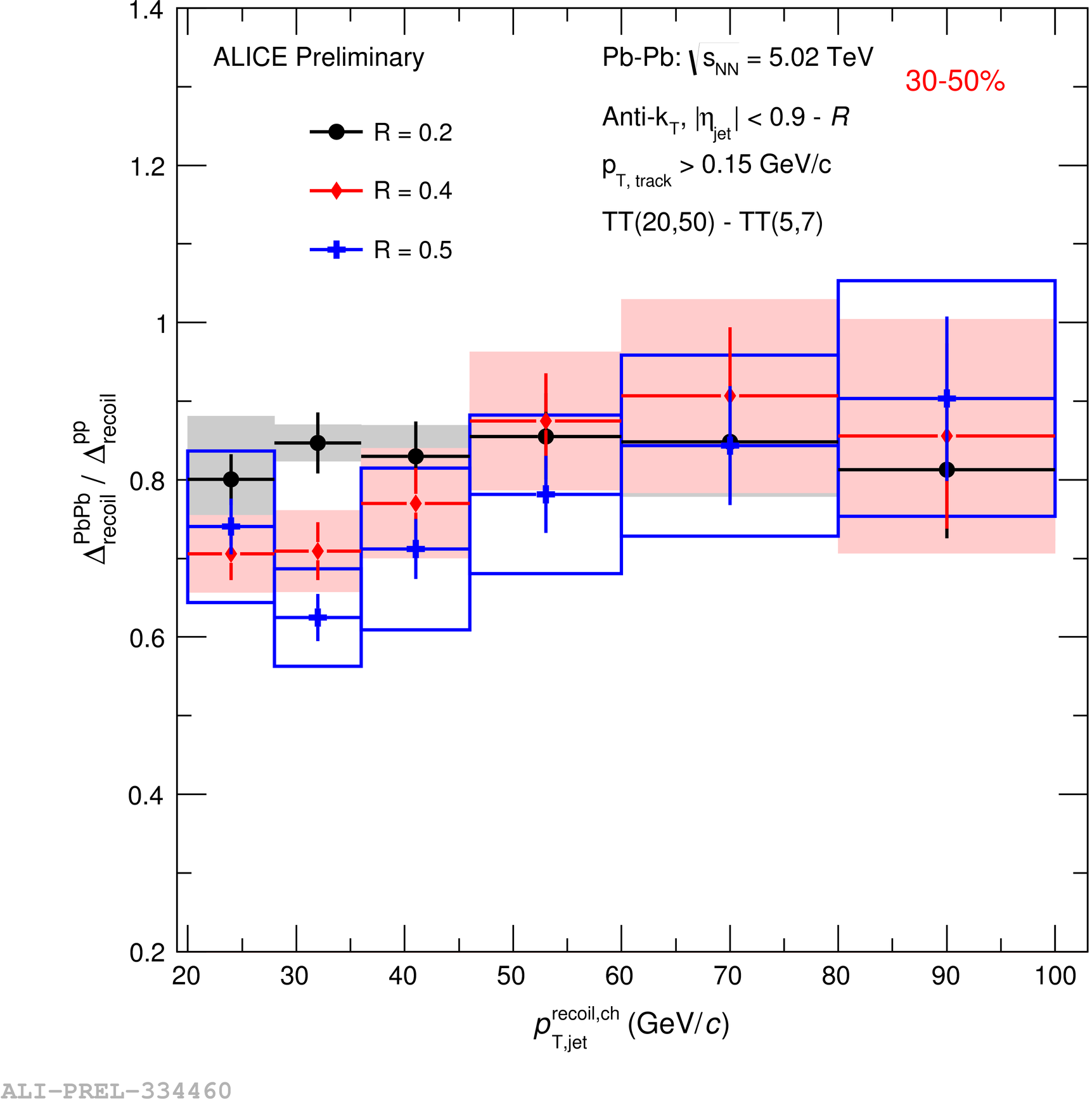}	
	\includegraphics[width=0.325\textwidth]{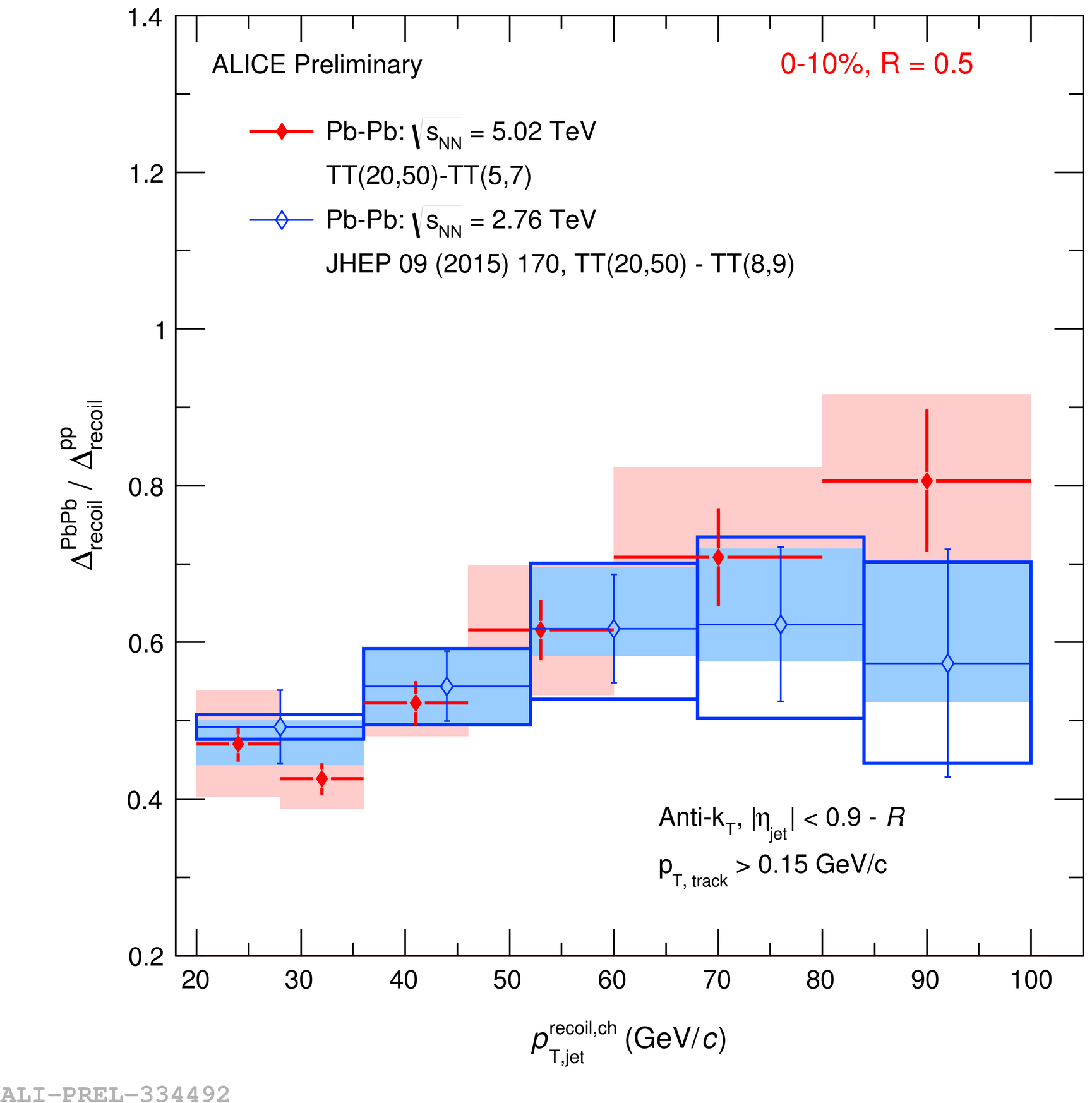}	
	\end{center}
	\caption[width=0.7\textwidth]{$\Delta_\mathrm{recoil}$ ratio in most central (0--10\%) and semi-central (30--50\%) Pb--Pb collisions to pp reference for $R$ = 0.2, 0.4 and 0.5, and comparisons between two different collision energies. }
	\label{fig:IAARecoil}
\end{figure*}

To study the jet quenching effect, we have calculated the ratio of the $\Delta_\mathrm{recoil}$ distribution from Pb--Pb collisions and pp collisions. Figure~\ref{fig:IAARecoil} shows the recoil jet yield suppressions in most central (0--10\%, left) and middle-central (30--50\%, middle) Pb--Pb collisions with respect to the same measurements in pp. A strong suppression of recoil jet yield is observed, especially in most central collisions. Such a suppression pattern is quite similar for different jet resolution parameters.  The right panel of Fig.~\ref{fig:IAARecoil} compares the $\Delta_\mathrm{recoil}$ ratio between two different collision energies. The results are consistent despite the different TT intervals and PYTHIA reference were used in the $\sqrt{s_{\mathrm {NN}}} = 2.76$ TeV analysis~\cite{ALICE_hJet}. 

\begin{figure*}[htbp]
	\begin{center}
	\includegraphics[width=0.325\textwidth]{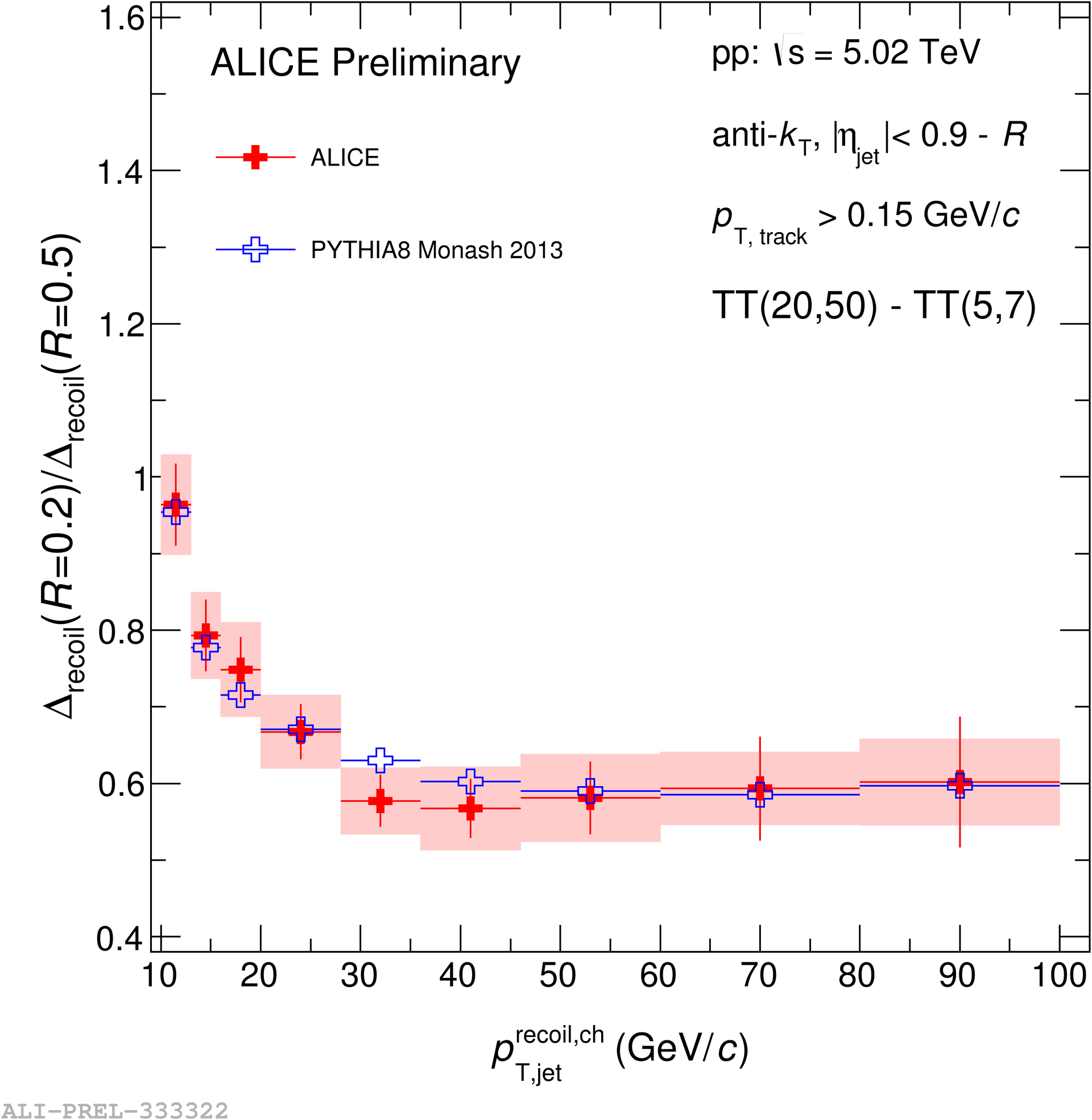}	
	\includegraphics[width=0.33\textwidth]{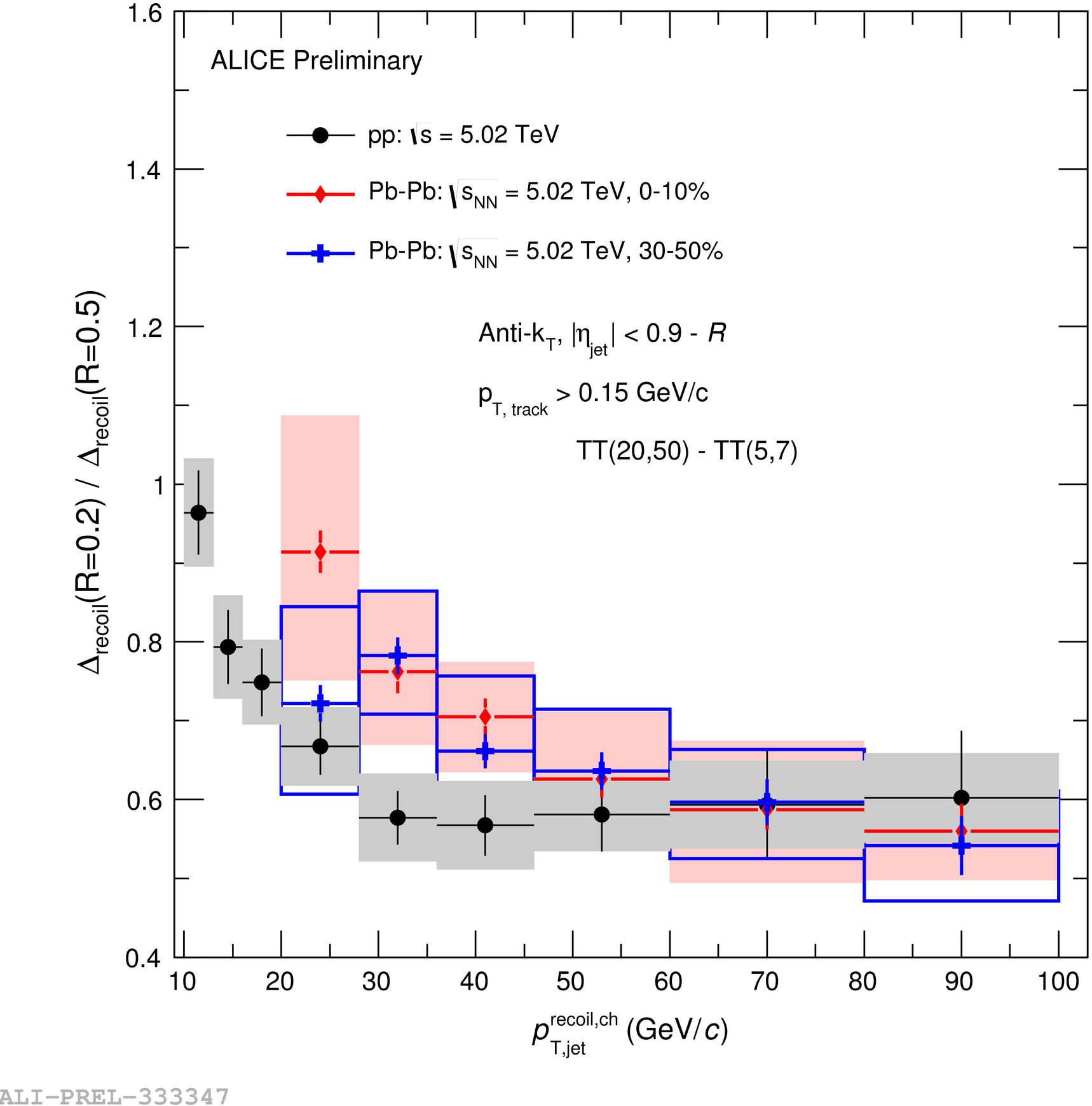}	
	\includegraphics[width=0.33\textwidth]{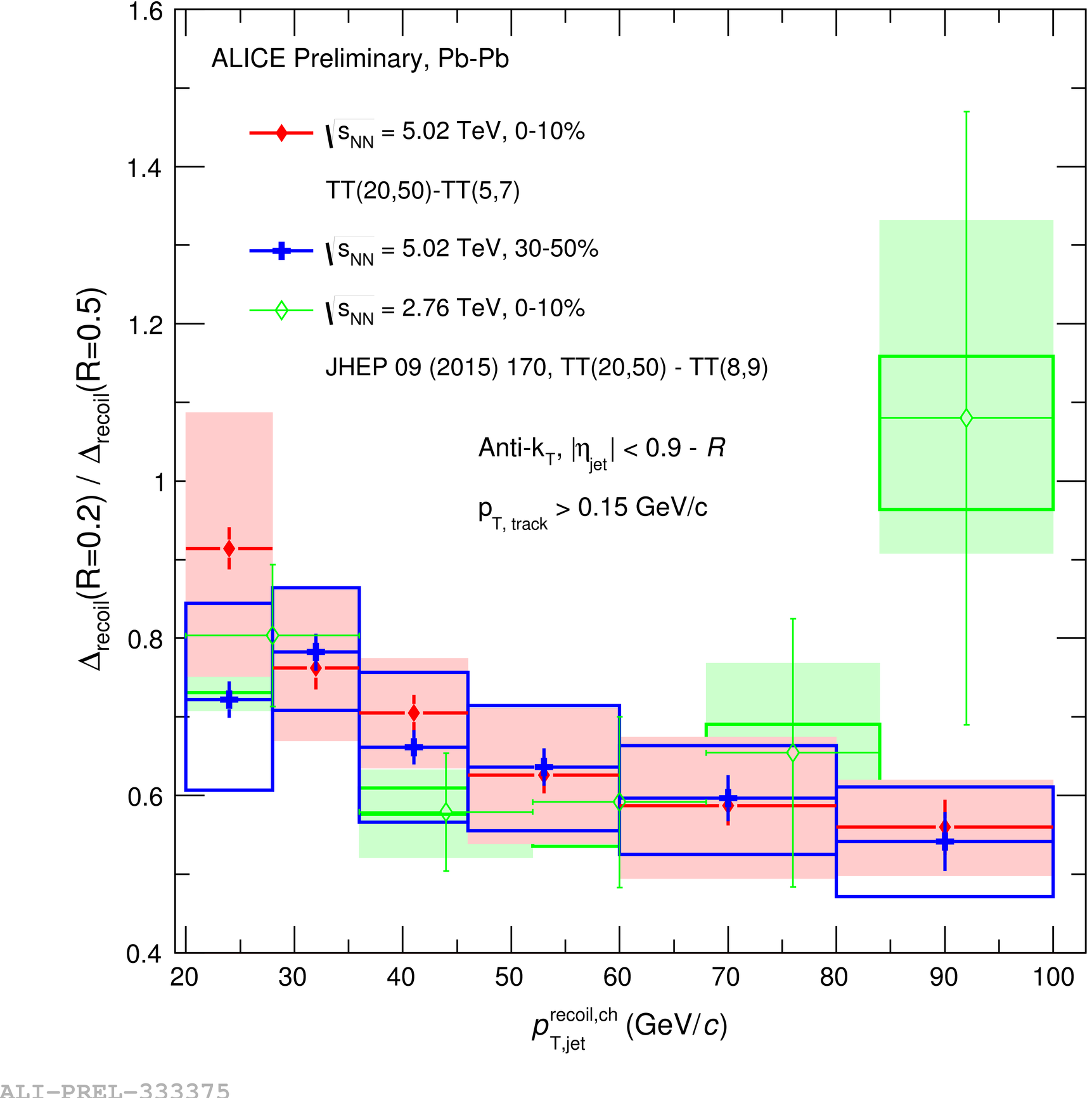}	
	\end{center}
	\caption[width=0.7\textwidth]{$\Delta_\mathrm{recoil}$ distribution ratio for $R$ = 0.2 to $R$ = 0.5 in pp collisions and PYTHIA MC (left) and Pb--Pb collisions using TT(20, 50) and TT(5,7) intervals. }
	\label{fig:RecoilJetRatio}
\end{figure*}

Figure~\ref{fig:RecoilJetRatio} shows the ratio of recoil jet yield reconstructed with a resolution parameter $R$ = 0.2 to those with $R$ = 0.5 in pp (left) and Pb--Pb (middle) collisions at $\sqrt{s_{\mathrm {NN}}} = 5.02$ TeV. 
The ratio of jet yields with different R is sensitive to the redistribution of jet energy due to quenching.
The recoil jet yield ratio in pp collisions is well described by PYTHIA MC (Fig.~\ref{fig:RecoilJetRatio}, left). When comparing the same ratio in different centrality Pb--Pb collisions to pp collisions, they are consistent between pp and Pb--Pb for high-$p_{\mathrm{T}}$ jets, but show a hint of jet energy redistribution in Pb--Pb collisions with respect to pp for low-$p_{\mathrm{T}}$ jets (Fig.~\ref{fig:RecoilJetRatio}, middle). 
The right panel of Fig.~\ref{fig:RecoilJetRatio} show the results obtained from RUN1~\cite{ALICE_hJet} and RUN2, they are consistent to each other within uncertainties. From the comparison, we conclude that the RUN2 analysis at $\sqrt{s_{\mathrm {NN}}} = 5.02$ TeV is much more precise with reduced uncertainties. 
\section{Summary and Outlook}
We have reported measurements of jet quenching in pp and Pb--Pb collisions at $\sqrt{s_{\mathrm {NN}}} = 5.02$ TeV, using the semi-inclusive distribution of jets recoiling from a high-$p_{\mathrm{T}}$ trigger hadron. The differential recoil jet yield in central Pb--Pb collisions is suppressed relative to that in pp collisions. The ratio of differential recoil jet yields with different $R$ is
similar for Pb--Pb and pp collisions at high-$p_{\mathrm{T}}$. A hint of medium-induced modification of the intra-jet energy distribution for angles $R \leq0.5$ relative to the jet axis is observed for low-$p_{\mathrm{T}}$ jets in central Pb--Pb collisions. In the future, the width of the azimuthal distribution of recoil jets relative to the trigger axis will be measured for low-$p_{\mathrm{T}}$ jets, where the medium-induced acoplanarity broadening is expected~\cite{Dijet,Acoplanarity,Moliere}. These new measurements, when combined with theoretical calculations, will provide new insights to discriminate the jet quenching mechanisms.  





\bibliographystyle{elsarticle-num}

\begin{thebibliography}{}
\expandafter\ifx\csname url\endcsname\relax
  \def\url#1{\texttt{#1}}\fi
\expandafter\ifx\csname urlprefix\endcsname\relax\def\urlprefix{URL }\fi
\expandafter\ifx\csname href\endcsname\relax
  \def\href#1#2{#2} \def\path#1{#1}\fi

\end{thebibliography}


\begin{thebibliography}{00}

\bibitem{Bjorken} Bjorken, J. D., {\em FERMILAB-PUB-82-059-THY} {1982}.
\bibitem{PHENIX} PHENIX Collaboration (K. Adcox {\em et al.}), {\em Phys. Rev. Lett.}{\bf 88} (2002) {022301}.
\bibitem{STAR} STAR Collaboration (C. Adler {\em et al.}), {\em Phys. Rev. Lett.}{\bf 89} (2002) {202301}.
\bibitem{CMS_Raa2012} CMS Collaboration (S. Chatrchyan {\em et al.}), {\em Eur. Phys. J. C} {\bf 72} (2012) {1945} .
\bibitem{ATLAS} ATLAS Collaboration (G. Aad {\em et al.}), {\em Phys. Rev. Lett.}{\bf 105} (2010) {252303}.
\bibitem{ALICE} ALICE Collaboration (K. Aamodt {\em et al.}), {\em Phys. Lett. B} {\bf 696} (2011) {30}.
\bibitem{ppjetALICE} ALICE Collaboration (S. Acharya {\em et al.}), {\em Phys. Rev. D} {\bf 100} (2019) {092004}.
\bibitem{ALICE_hJet} ALICE Collaboration (J. Adam {\em et al.}), {\em JHEP} {\bf 09} (2015) {170}.
\bibitem{detector} ALICE Collaboration (K. Aamodt {\em et al.}), {\em JINST} {\bf 3} (2008) {S08002}.
\bibitem{Cacciari:2011ma} M.~Cacciari, G.~P. Salam, and G.~Soyez, {\em Eur. Phys. J. C} {\bf 72} (2012) {1896}.
\bibitem{FastJetAntikt} M.~Cacciari, G.~P. Salam, and G.~Soyez, {\em JHEP} {\bf 04} (2008) {063} .
\bibitem{deBarros:2012ws} G.~de~Barros, B.~Fenton-Olsen, P.~Jacobs, and M.~Ploskon, {\em Nucl.Phys. A} {\bf 910} (2013)  {314}.
\bibitem{DAgostini:1994zf} D'Agostini, G., {\em Nucl. Instrum. Meth. A} {\bf 362} (1995) {487}.
\bibitem{Pythia8} T. Sj\"ostrand, S. Mrenna and P. Skands, {\em Comput. Phys. Commun.}, {\bf 178} (2008) {852}.
\bibitem{Dijet} L. Chen, G. Qin, S. Wei, B. Xiao, and H. Zhang, {\em Phys. Lett. B} {\bf 782} (2018) {773}.
\bibitem{Acoplanarity} M. Gyulassy, P. Levai, J. Liao, S. Shi, F. Yuan and X. Wang, {\em Nucl. Phys. A} {\bf 982} (2019) {627}.
\bibitem{Moliere} F. D'Eramo, K. Rajagopal and Y. Yin, {\em JHEP} {\bf 01} (2019) {172}.

 \end{thebibliography}



\end{document}